\documentclass[]{aastex}

\usepackage{emulateapj5}
\usepackage{onecolfloat}
\usepackage{graphicx} 
\usepackage{fancyheadings} 
\usepackage{ulem}
\usepackage{rotating}
\usepackage{lscape}

\newcommand{\odla}{$\Omega_g^{\rm DLA}$}
\newcommand{\oprox}{$\Omega_g^{\rm PDLA}$}

\newcommand{\npdla}{111}
\newcommand{\kms}{km~s$^{-1}$}
\newcommand{\cm}[1]{\, {\rm cm^{#1}}}
\newcommand{\mkms}{{\rm \; km\;s^{-1}}}

\newcommand{\lya}{Ly$\alpha$}
\newcommand{\lyb}{Ly$\beta$}

\newcommand{\sci}[1]{{\rm \; \times \; 10^{#1}}}

\newcommand{\mnhi}{N_{\rm HI}}

\newcommand{\nhi}{$N_{\rm HI}$}
\newcommand{\fnhi}{$f(N_{\rm HI},X)$}
\newcommand{\mfnhi}{f(N_{\rm HI},X)}

\def\smm{\sum\limits}
\def\intl{\int\limits}
\def\loz{$\ell(z)$}
\def\mloz{\ell(z)}
\def\lox{$\ell(X)$}
\def\lpdla{$\ell(X)_{\rm PDLA}$}

\def\ltp{\left ( \,}

\def\rtp{\, \right  ) }

\def\nhi{$N_{\rm HI}$}
\def \hMpc      {h^{-1}{\rm\ Mpc}}

\def\perd{\;\;\; .}
\def\cmma{\;\;\; ,}

\begin{document}
\twocolumn[%
\submitted{Submitted to ApJ; March 16 2007}

\title{The SDSS-DR5 Survey for Proximate Damped \lya\ Systems}

\author{Jason X. Prochaska\altaffilmark{1}, Joseph
  F. Hennawi\altaffilmark{2,3}, St\'ephane
  Herbert-Fort\altaffilmark{4}}

\begin{abstract}
Using the Sloan Digital Sky Survey, Data Release 5 (SDSS-DR5), we
survey proximate damped \lya\ systems (PDLAs): absorption line systems
with \ion{H}{1} column density $\mnhi \ge 2 \sci{20} \cm{-2}$ at
velocity separation $\delta v < 3000 \mkms$ from their background
quasar.  These absorbers are physically associated with their
background quasars, and their statistics allow us to study quasar
environments out to $z\sim 5$.  However, the large ionizing flux
emitted by a quasar can ionize the neutral gas in a nearby galaxy
possibly giving rise to a ``proximity effect'', analogous to the
similar effect observed in the \lya\ forest.  From a sample of 
\npdla, we measure the
\ion{H}{1} frequency distribution \fnhi, incidence, and gas mass
density of the PDLAs near luminous quasars over the redshift interval
$z= 2.2$ to 5.  The incidence and mass density of PDLAs at $z\sim 3$
is approximately twice that of intervening DLAs, but at $z<2.5$ and
$z>3.5$ the \fnhi\ distribution is consistent with the
intervening population.  We interpret the observed enhancement of
PDLAs around quasars in terms of quasar-galaxy clustering, and compare
the strength of the clustering signal to the expectation from
independent measures of the respective clustering strengths of DLAs
and quasars, as well as a complementary analysis of the clustering of
absorbers around quasars in the transverse direction. We find that
there are a factor of $5-10$ fewer PDLAs around quasars than expected
and interpret this result as evidence for the hypothesis that the ionizing
flux from the quasars photoevaporates \ion{H}{1} in nearby
DLA galaxies, thus reducing their cross-section for DLA
absorption. This constitues the first detection of a ``proximity
effect'' for DLAs.  

\altaffiltext{1}{Department of Astronomy and Astrophysics, 
UCO/Lick Observatory;
University of California, 1156 High Street, 
Santa Cruz, CA 95064; xavier@ucolick.org}
\altaffiltext{2}{Department of Astronomy, 601 Campbell Hall, 
        University of California, Berkeley, CA 94720-3411}
\altaffiltext{3}{Hubble Fellow}
\altaffiltext{4}{University of Arizona/Steward Observatory, 
933 N Cherry Avenue, Tucson, AZ 85721}

\end{abstract}
]

\section{Introduction}

Although quasars are believed to reside at the centers of massive
galaxies which are themselves located in groups or clusters, their
spectra rarely exhibit the absorption signature of either their host's
neutral interstellar medium (ISM) or the ISMs of associated nearby
galaxies\footnote{The relatively few cases where one observes
  significant absorption at $z \approx z_{em}$ are characterized by
  very broad features and are termed Broad Absorption Line (BAL)
  quasars.  These BAL quasars are typically attributed to gas with
  extreme densities and large velocity fields residing in the broad
  line region of the quasar
  \citep[e.g.][]{weymann91,tolea02,reich03,trump+06} and should not be
  confused with the much larger galactic scale absorption which causes
  DLAs.}.  Nearly every galaxy at $z \approx 3$ should exhibit an ISM
comprised of neutral hydrogen.  If the quasar sightline penetrates this
ISM, it will absorb light along the hydrogen Lyman series, with
strongest absorption at \lya: $\lambda_{\rm rest} = 1215.67$\AA.
Systems with \ion{H}{1} column densities \nhi\ exceeding $2 \sci{20}
\cm{-2}$ are termed damped \lya\ (DLA) systems owing to the strong
damping wings that they exhibit \citep[see][for a review]{wgp05}. Gas
clouds at these high column densities, characteristic of a galactic
disk, are optically thick to Lyman continuum ($\tau_{\rm LL} \gg 1$) 
photons, giving rise to a neutral interior self-shielded from the
extragalactic ionizing background.


One might guess that the absence of DLAs near most quasars is related
to the large ionizing flux emitted by the quasar.  In particular, for
a quasar at $z=3$ with an $i$-band magnitude of $i=19.1$, the flux of
ionizing photons is 400 times higher than that of the extragalactic UV
background at a comoving distance of $1~\hMpc$ (corresponding to
Hubble flow velocity of $100$~\kms), and increasing as $r^{-2}$ toward
the quasar. Indeed, the decrease in the number of optically
thin absorption lines ($\log N_{\rm HI}<17.2$ hence $\tau_{\rm
  LL}\lesssim 1$), in the vicinity of quasars, known as the
\emph{proximity effect} \citep{bdo88}, is well studied and has been
detected \citep{sbd+00,claude07}.

On the other hand, it has long been known that quasars are associated
with enhancements in the distribution of galaxies
\citep{BSG69,YG84,YG87,BC91,SBM00,BBW01,Serber06,coil06}, although
these measurements of quasar galaxy clustering are mostly limited to
low redshifts $\lesssim 1.0$. Recently, \citet{as05}, measured the
clustering of Lyman Break Galaxies (LBGs) around luminous quasars in
the redshift range ($2 \lesssim z \lesssim 3.5$), and found a best fit
correlation length of $r_0=4.7~\hMpc$ ($\gamma=1.6$), very similar to
the auto-correlation length of $z\sim 2-3$ LBGs
\citep{adel03,adel05}. \citet{cwg+06-2} recently measured the
clustering of LBGs around DLAs and measured a best fit $r_0=2.9~\hMpc$
with $\gamma=1.6$, but with large uncertainties \citep[see
also][]{Gawiser01,Bouche04}.  If LBGs are clustered around quasars,
and DLAs, might we expect DLAs to be clustered around quasars?  This
is especially plausible in light of recent evidence that DLAs arise
from a high redshift galaxy population which are not unlike LBGs
\citep{mwf+02}.

Can DLAs continue to self-shield against the large flux of ionizing
photons emitted by quasars or are they be photoevaporated? Is the
distribution of DLAs around quasars dominated by ionization effects or
does galaxy clustering around the quasars dominate?  Does the column
density distribution of DLAs near quasars differ from the distribution
in average places in the Universe?  What can quasar-DLA clustering
teach us about quasars and the physical nature of high redshift
galaxies?

In this paper, we take positive step towards answering these questions
by measuring the abundance and distribution of DLAs near quasars.
Recently, \cite[][; hereafter PHW05]{phw05} surveyed the quasar spectra
of the Sloan Digital Sky Survey (SDSS) Data Release 3 \citep{sdssdr3} for
damped \lya\ systems at $z>2.2$.  The statistical survey, comprising
over 500 DLA systems, provides a relatively precise measure of the
incidence of intervening DLA systems per unit redshift, \loz, from
$z=2.2$ to 4.  To focus on the cosmic average of \loz and avoid biases
from the quasar environment, PHW05 excluded \emph{proximate DLAs}
(PDLAs) from their analysis, which are defined to be DLAs with
velocity offset $\delta v < 3000 \mkms$ from the emission redshift of
the background quasars\footnote{At $z=3, \delta v = 3000 \mkms$
  corresponds to roughly 30 comoving Mpc~$h^{-1}$ assuming 
$\Omega_m = 0.3, \Omega_\Lambda = 0.7$ and 
$H_0 = 70$\,km\,s$^{-1}$\,Mpc$^{-1}$.}.  
Here we present the results of a survey
for proximate damped \lya\ (PDLA) systems in the full SDSS Data
Release 5 \citep[DR5][]{sdssdr5}.

The first statistical study of PDLAs was performed by \cite{eyh+02}
who surveyed their sample of radio-selected QSOs.
Our work is motivated by this and several recent, closely related
studies. First, Russell et al.\ (2006; hereafter REB06) have previously
performed a search of a subset of the SDSS-DR3 quasar spectra for
PDLAs.  They reported an enhancement of \loz\ for DLA systems with
$\delta v < 3000 \mkms$ and, oddly, an enhancement of PDLAs with $3000
\mkms < \delta v < 6000 \mkms$, corresponding to very large distances
($30-60~\hMpc$) from the quasar if interpreted as Hubble flow.  Our
search builds on their work, but we survey the larger SDSS-DR5 to
recover a final sample with nearly an order of magnitude more PDLA
systems.  Second, \citet{hpb+06} published a large sample of optically thick
absorption line systems in the vicinity of $z\sim 2.5$ quasars, which
were identified using a background line-of-sight in close projected
quasar pair systems. Based on this sample, \citet{hp06} measured the
cross-correlation function between foreground quasars at $z\sim 2.5$
and optically thick $\mnhi > 10^{19}~{\rm cm}^2$ absorbers, detected
in the background quasar spectra. Their measurement of the
\emph{transverse} clustering of absorbers around quasars should be
commensurate with incidence of similar systems along the
line-of-sight, provided that the clustering pattern of absorbers
around quasars is isotropic. However, \citet{hp06} argued that the
strength of the transverse clustering predicts that $\sim 15-50\%$ of
\emph{all} quasars should show a $\mnhi > 10^{19} \cm{-2}$ absorber
within $\delta v < 3000$~\kms, which is certainly not observed.
Finally, another goal of our survey is to identify candidates for
follow-up observations to study \lya\ emission from PDLAs which could
possibly yield insights into fluorescent \lya\ recombination radiation
from DLAs, the size of DLA galaxies, and the physics of \lya\ halos
surrounding high $z$ quasars \citep{joe07}.

In the following section, we describe the survey and search for PDLA
candidates and, in $\S$~\ref{sec:nhi}, we discuss their \nhi\
measurements.  A new estimate of the quasar redshifts with PDLA
candidates is given in $\S$~\ref{sec:zem}.  The \ion{H}{1} frequency
distribution and incidence of PDLA systems is presented in
$\S$~\ref{sec:fnhi} and we discuss these results in terms of
clustering and the proximity effect in $\S$~\ref{sec:discuss}. We summarize
and conclude in $\S$~\ref{sec:summary}.
Throughout the paper we adopt a $\Lambda$CDM cosmology with 
$\Omega_m = 0.3, \Omega_\Lambda = 0.7$ and $H_0 = 70 \mkms$\,Mpc$^{-1}$.

\section{Survey Design and PDLA Candidates}
\label{sec:survey}

We began with the full SDSS-DR5 sample of $\sim$67000
spectroscopically classified quasars.  
The SDSS fiber-fed spectrometer gives a FWHM resolution of 
$\approx 2$\AA\ and wavelength coverage spanning 3800 to 9200\AA.
Following PHW05, we demanded that the quasar spectrum
exhibit at least one region where the median signal-to-noise (S/N) of 20
consecutive pixels equals or exceeds 4. 
This requirement eliminates many of the quasars at $z>4$
which have poorer S/N data.  For each QSO,
we determine a starting and ending redshift $z_i$ and $z_f$
which determines the search path of that object:

\begin{equation}
(\Delta z)_j \equiv (z_i - z_f)_j \perd
\label{eqn:deltaz}
\end{equation}
In the PHW05 analysis, $z_i$ corresponded to the first pixel
where the data satisfied the S/N requirement under the
restriction that this not occur within 10000\kms\ of the
\lyb+\ion{O}{6} emission region of the QSO\footnote{Note
that PHW05 failed to implement this restriction
in their published sample.  Although the differences to the
computed statistics of the intervening DLA systems are small,
future analyses by those authors
will properly implement the algorithm and the authors provide
the correct SDSS-DR3 statistical
sample at http://www.ucolick.org/$\sim$xavier/SDSSDLA.
The analysis here includes the proper statistical sample.
\label{foot:ovi}}.
For a search focusing on PDLA systems, we will define 
$z_i$ to be a velocity $v_{prox}$ blueward of $z_{em}$ under a strict
restriction:
the quasars must also satisfy the S/N criterion at a velocity
greater than (3000\kms\ + $v_{prox}$) from the QSO. 
We adopt this restriction for two
reasons: (i) we wish to avoid the presence of a DLA system
with $v \approx v_{prox}$ from biasing the S/N below our criterion;
(ii) as discussed in $\S$~\ref{sec:fnhi}, we wish to maintain
a search path $(\Delta z)_j$ for each QSO to be set exactly by $v_{prox}$.

\begin{table*}[ht]\footnotesize
\begin{center}
\caption{{\sc SDSS-DR5 QUASAR SAMPLE FOR PDLA\label{tab:qso}}}
\begin{tabular}{ccccccccl}
\tableline
\tableline
Plate &MJD  & FiberID & Name  & $z_{qso}^a$ & $z_{new}^b$ & $f_{BAL}^c$ & $z_{i}^d$ & $z_{candidate}$ \\
\tableline
 750&52235&  82&J000009.38$+13$5618.4&2.240&...&0&... &\\
 650&52143& 178&J000050.60$-10$2155.8&2.640&...&0&2.604&\\
 750&52235& 550&J000143.41$+15$2021.4&2.638&...&0&2.602&\\
 650&52143& 519&J000159.12$-09$4712.4&2.308&...&0&2.276&\\
 387&51791& 556&J000221.11$+00$2149.4&3.057&...&0&3.017&\\
 650&52143& 111&J000238.41$-10$1149.8&3.941&...&0&... &\\
 750&52235& 608&J000300.34$+16$0027.7&3.675&...&0&3.629&\\
 650&52143&  48&J000303.34$-10$5150.6&3.647&...&1&3.600&\\
 750&52235&  36&J000335.21$+14$4743.6&3.484&...&0&... &\\
 650&52143& 604&J000413.63$-08$5529.5&2.424&...&0&2.390&\\
\tableline
\end{tabular}
\end{center}
\tablenotetext{a}{Quasar redshift reported in SDSS-DR5.}
\tablenotetext{b}{New quasar redshift estimated using the algorithms developed by \cite{hpb+06} and \cite{shen07}.  This analysis was 
performed on quasars with PDLA candidates lying within 10,000\kms
of the SDSS-DR5 redshift.}
\tablenotetext{c}{0=No intrinsic absorption; 1=Mild intrinsic absorption, included in analysis; 2=Strong intrinsic absorption,  excluded.}
\tablenotetext{d}{Starting redshift for the PDLA search assuming
an offset of $v_{prox} = 3000\mkms$ from the quasar, allowing
for the revised redshifts in column~6.}
\tablecomments{[The complete version of this table is in the electronic edition of the Journal.  The printed edition contains only a sample.]}
\end{table*}

\begin{figure}[ht]
\begin{center}
\includegraphics[height=3.5in,angle=90]{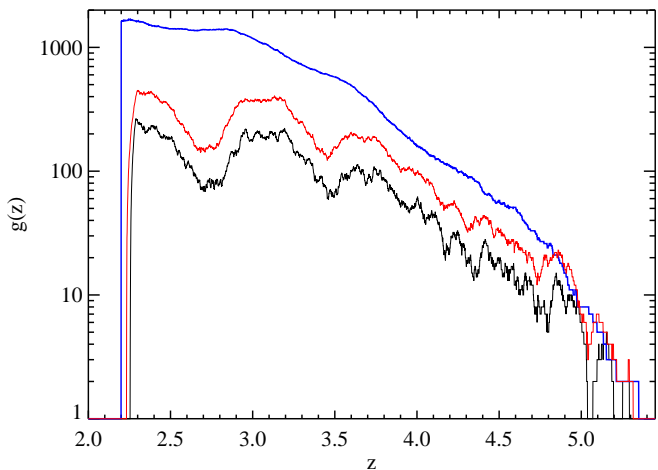}
\end{center}
\caption{Curves of the redshift selection function $g(z)$ as a function of
redshift for (top) quasars satisfying the search criterion of PHW05 for
DLA drawn from the SDSS-DR3; 
(middle) quasars drawn from the SDSS-DR5 for a search within
$v_{prox}=6000$ of the quasar emission redshift $z_{em}$; and
(bottom)  quasars drawn from the SDSS-DR5 for a search within
$v_{prox}=3000$ of $z_{em}$.
The bin size in the figure is $\delta z = 10^{-3}$.
}
\label{fig:gofz}
\end{figure}

The total redshift search path of the survey can be summarized
by its sensitivity function \citep[e.g.][]{lzt93}:

\begin{equation}
g(z) = \smm_j \Theta(z)_j
\label{eqn:goz}
\end{equation}
where $\Theta(z)_j$ is unity if $z + \delta z$ is included
within a given quasar's search path $(\Delta z)_j$ 
and zero otherwise.  The $g(z)$ function 
for the SDSS-DR5 database is shown in 
Figure~\ref{fig:gofz} for $v_{prox} = 3000$ and 6000\kms\ 
(bottom and middle curves respectively).
For comparison, we show the same quantity for the intervening
DLA search of PHW05 (corrected for the error described in 
footnote~\ref{foot:ovi}).  In contrast to the intervening $g(z)$
curve, the PDLA $g(z)$ curves show significantly more structure.
The $g(z)$ curves closely track the redshift distribution of
SDSS quasars and most of the structure is the result of the
SDSS quasar target selection algorithm \citep{rng+04}.  
On average, the intervening DLA search path is $\approx 10\times$
higher than the PDLA path for $z<3.5$ and $\approx 5\times$
higher at $z \approx 4$.  If the incidence of PDLA systems
matches that of intervening DLAs, then we will expect
to find a statistical sample with $5$ to $10\times$ fewer systems than
PHW05, i.e.\ $\approx 75$ systems.

Every quasar with an 
emission redshift $z_{em}$ measured to be greater than 2.2 by
the SDSS pipeline was searched for a PDLA candidate.  The
search algorithm is nearly identical to the algorithm described
in PHW05 for intervening absorbers, only modified to find
candidates up to 7000\kms\ redward of $(1+z_{em}) \times 1215.67$.
The algorithm keys on any regions in the spectrum
where the S/N is coherently low over a window of $\approx 25$\AA.
The algorithm triggers on the cores of damped \lya\ systems,
but also on the metal-line absorption features of strong BAL
systems.  Therefore, we 
removed BAL quasars from the search sample;  all of the SDSS-DR5
quasars were visually inspected and those exhibiting very strong
\ion{C}{4} profiles at $z \approx z_{em}$ were eliminated.
Because many PDLA systems will exhibit strong \ion{C}{4} absorption,
we chose to flag only those where the \ion{C}{4} doublet
is blended with itself (i.e. $\Delta v_{CIV} > 500 \mkms$) and
the rest equivalent width $W_\lambda$ exceeds $\approx 3$\AA.
We admit, however, that this `by-eye' analysis is somewhat
subjective and, in general, we conservatively included 
border-line cases under the expectation that a true BAL would
not show damped \lya\ absorption.
All of the quasars comprising the search are listed in 
Table~\ref{tab:qso}.

\begin{figure}[ht]
\begin{center}
\includegraphics[height=3.5in,angle=90]{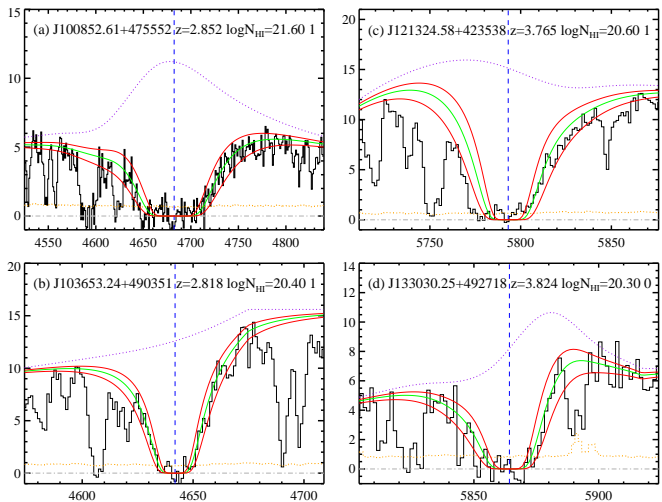}
\end{center}
\caption{A set of representative fits from our analysis of PDLA candidates.
The dotted line above the fit
traces the predicted continuum of the quasar.  It is 
complicated by the presence of \lya\ and \ion{N}{5} emission.
The fit is shown by the light solid line and the estimated $1\sigma$
uncertainty is given by the darker lines which bracket it.
The full set of fits are presented at
http://www.ucolick.org/$\sim$xavier/SDSSDLA.
}
\label{fig:dlaex}
\end{figure}

\section{\nhi\ MEASUREMENTS}
\label{sec:nhi}

Our procedure to measure the \nhi\ values of the PDLA candidates
follows that of PHW05 and, in fact, we adopt their published
values for PDLA candidates from the Data Release 3.
For each new PDLA candidate, we searched for the presence of 
metal-line absorption near the redshift of the \lya\ centroid.
Because the absorption redshift coincides with the quasar
emission redshift ($z_{abs} \approx z_{em}$), the strong low-ion transitions
of \ion{Si}{2}~1206, \ion{C}{2}~1334, and \ion{Al}{2}~1670
generally lay clear of the \lya\ forest and we found that nearly
every PDLA system exhibit at least one of these transitions.
We defined $z_{abs}$ to be the centroid of the strongest, unblended 
low-ion transition.

We then fit a Voigt profile to the candidate's \lya\
profile while also estimating the quasar continuum.  The
latter effort is significantly complicated by the quasar's
\lya\ and \ion{N}{5} emission lines.   We guided the continuum
placement by the observed heights of the \lyb\ and \ion{C}{4}
emission features, but the systematic uncertainty related to this
exercise dominates the errors reflected in our measurements of 
\nhi.  Therefore, while we have reported relatively conservative
errors for the measurements, we are certain the error distribution
is non-Gaussian.  For example, we expect that the probability of an
error greater than $3 \sigma$ is not less than $0.27\%$.  
We do believe, however, that $\approx 95\%$ of the values
lie within $2\sigma$ of the reported \nhi\ measurements.

\begin{table}[ht]\footnotesize
\begin{center}
\caption{{\sc SDSS-DR5 PROXIMATE DLA STATISTICAL SAMPLE\label{tab:dlastat}}}
\begin{tabular}{lccc}
\tableline
\tableline
Quasar & $z_{abs}$ & $f_{mtl}^b$ & log \nhi \\
\tableline
J004142.5$-08$5704.6&3.6069&2&$20.65^{+0.20}_{-0.20}$\\
J014049.1$-08$3942.5&3.6966&2&$20.70^{+0.15}_{-0.15}$\\
J014214.7$+00$2324.3&3.3482&2&$20.40^{+0.15}_{-0.15}$\\
J074823.8$+33$2051.2&2.9755&2&$20.45^{+0.15}_{-0.15}$\\
J075901.2$+28$4703.4&2.8226&2&$21.05^{+0.15}_{-0.15}$\\
J080523.3$+21$4921.1&3.4774&0&$21.65^{+0.15}_{-0.15}$\\
J080553.0$+30$2937.3&3.4294&2&$20.40^{+0.15}_{-0.15}$\\
\tableline
\end{tabular}
\end{center}
\tablenotetext{a}{0=No metals; 1=Weak metals; 2=Metals detected}
\tablecomments{[The complete version of this table is in the electronic edition \\
 of the Journal.  The printed edition contains \\
only a sample.]}
\end{table}
 
\begin{table}[ht]\footnotesize
\begin{center}
\caption{\sc {SDSS-DR5 PROXIMATE DLA NON-STATISTICAL SAMPLE\label{tab:dlanon}}}
\begin{tabular}{lccc}
\tableline
\tableline
Quasar & $z_{abs}$ & $f_{mtl}^b$ & log \nhi \\
\tableline
J001134.5$+15$5137.4&4.3175&1&$20.50^{+0.20}_{-0.20}$\\
J001134.5$+15$5137.4&4.3592&2&$21.10^{+0.20}_{-0.20}$\\
J014214.7$+00$2324.3&3.3482&2&$20.40^{+0.15}_{-0.15}$\\
J074823.8$+33$2051.2&2.9755&2&$20.45^{+0.15}_{-0.15}$\\
J075901.2$+28$4703.4&2.8226&2&$21.05^{+0.15}_{-0.15}$\\
J080553.0$+30$2937.3&3.4294&2&$20.40^{+0.15}_{-0.15}$\\
J081114.3$+39$3633.2&3.0415&1&$20.85^{+0.20}_{-0.20}$\\
J081543.1$+37$0037.0&3.1780&2&$20.30^{+0.20}_{-0.20}$\\
J081813.0$+26$3136.9&4.1610&2&$21.10^{+0.15}_{-0.15}$\\
J082021.3$+39$0327.2&4.2838&2&$20.45^{+0.15}_{-0.15}$\\
\tableline
\end{tabular}
\end{center}
\tablenotetext{a}{0=No metals; 1=Weak metals; 2=Metals detected}
\tablecomments{[The complete version of this table is in the electronic edition \\
 of the Journal.  The printed edition contains \\
only a sample.]}
\end{table}

A set of representative fits is shown in Figure~\ref{fig:dlaex}.
One notes, in several cases, the challenges of continuum placement
(dotted line) near the \lya\ emission line.  One also notes cases where
the PDLA \lya\ profile is significantly blueward of the quasars
\lya\ emission line.  This is a clear indication that 
determinations of $z_{em}$ based primarily on \lya\ emission will 
give a poor result.  We will return to this issue in the following
section.  The complete set of fits can be found
at http://www.ucolick.org/$\sim$xavier/SDSSDLA.
Finally, we wish to briefly compare the \nhi\ values we derived
with those reported in REB06 (measured from the same
spectra).  As REB06 noted, the agreement between their
vales and PHW05 is good.  We identify a few cases
(e.g.\ the DLA at $z=3.958$ toward J111224.18+004630.3), however, 
where we find much lower \nhi\ values than REB06.  In these
cases, we have verified our lower values by analyzing 
the corresponding \lyb\ profile.

\begin{figure}[ht]
\plotone{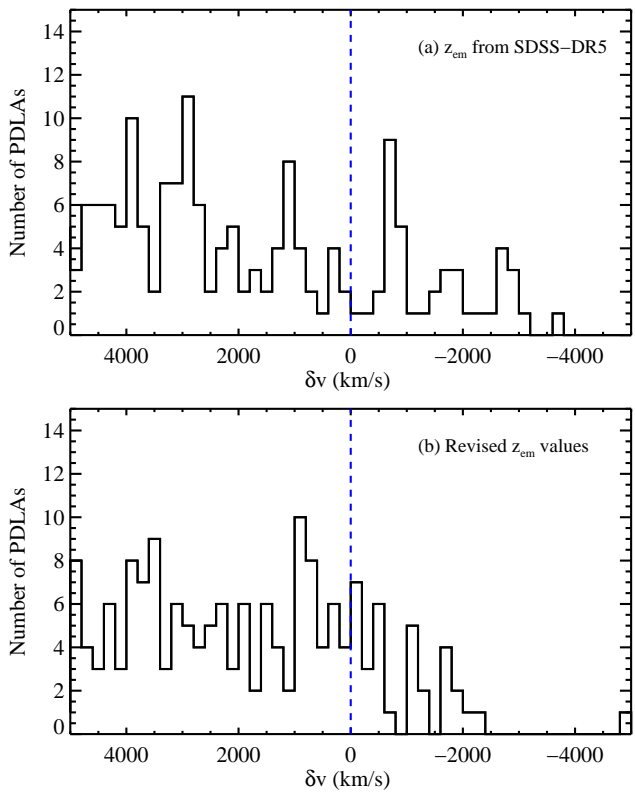}
\caption{(a) Velocity offset $\delta v$
between $z_{abs}$ of the PDLA candidates and the 
quasar emission redshift $z_{em}$ reported in SDSS-DR5.
Negative velocities indicate $z_{abs} > z_{em}$.
This analysis demonstrates that there are significant
systematic errors in the emission redshifts; note, especially,
that there are few examples with $\delta v \approx 0$
yet many PDLA candidates with $\delta v < -500 \mkms$.
These results motivated us to re-measure the redshifts of
all quasars with a PDLA candidate within 10,000\kms\ using
the techniques of \cite{hpb+06} and \cite{shen07}.
(b) Same as (a) but adopting new quasar redshifts after employing
new algorithms for the quasar emission redshift
\citep{hpb+06,shen07}.
The new $\delta v$ values have a more physical 
distribution near 0\kms.
}

\label{fig:histvel}
\end{figure}

\section{NEW MEASUREMENTS OF QUASAR REDSHIFTS}
\label{sec:zem}

In Figure~\ref{fig:histvel}a we present a histogram of
the velocity offsets, 

\begin{equation}
\delta v \equiv c \frac{R^2 - 1}{R^2 + 1} \cmma
\label{eqn:dv}
\end{equation}
with $R \equiv (1+z_{abs})/(1+z_{em})$ between $z_{abs}$ for all of the
fitted PDLA candidates and $z_{em}$ from its corresponding quasar as
listed in the SDSS-DR5 database.  It is evident from the figure
that the quasar redshifts are systematically in error.  
The presence of very few PDLA candidates with 
$\delta v \approx 0 \mkms$ and many examples with very negative $\delta v$
suggests that the \lya\
absorption profiles of PDLA systems have significantly biased
the measurement of $z_{em}$ away from its correct value. 

It is well known that the primary rest-frame ultraviolet quasar
emission lines which are redshifted into the optical for $z\gtrsim 2$,
can differ by up to $\sim 3000$~\kms\ from systemic, due to
outflowing/inflowing material in the broad line regions of quasars
\citep{gaskell82,TF92,vanden01,richards02}. The quasar redshifts
provided by the SDSS spectroscopic pipeline come from a maximum
likelihood fit to multiple emission lines \citep[see
e.g. ][]{Stoughton02}, which does not result in robust systemic
redshifts estimates. Concerned that the errors in $z_{em}$ will
significantly affect our clustering analysis, we estimated the
systemic emission redshift of every QSO with a PDLA candidate
exhibiting $\mnhi > 10^{20} \cm{-2}$ and $\delta v < 10000 \mkms$,
following the approach described in \citet[][see also Hennawi et
al. 2006]{shen07}.  \citet{shen07} measured the correlation between
the relative shifts of the high-ionization emission lines \ion{Si}{4},
\ion{C}{4}, \ion{C}{3}, and the shift between these respective lines
and the \ion{Mg}{2} line.  Since the redshift defined by \ion{Mg}{2}
is tightly correlated with the systemic redshift \citep{richards02},
the \citet{shen07} approach exploits these correlations to ``shift''
into the systemic frame. The emission lines in the SDSS spectra were
centered using the algorithm described in \citet{hpb+06} which was
also employed by \citet{shen07}. The updated redshifts are presented
in column~6 of Table~\ref{tab:qso}.

In the lower panel of Figure~\ref{fig:histvel}, we present
the velocity offsets for all PDLA candidates with $\delta v < 10000 \mkms$.
We note a nearly smooth distribution of systems with $\delta v > 0 \mkms$
and $\approx 40$ PDLA with $\delta v < 0 \mkms$.  Although we
expect the revised redshifts still suffer from systematic uncertainties,
we are optimistic that the uncertainty is smaller than a few
hundred \kms\ and, more importantly, that the errors
are symmetrically distributed.
One may be concerned that we have restricted the new $z_{em}$
measurements to a subset of the SDSS-DR5 QSO database.
As discussed below, because we include all PDLA systems with
negative $\delta v$ in our clustering analysis, the error in
$z_{em}$ for quasars without PDLA candidates should have minimal
effect on our results.

\begin{table*}[ht]\footnotesize
\begin{center}
\caption{\sc {FITS TO \fnhi\label{tab:fnfits}}}
\begin{tabular}{ccccc}
\tableline
\tableline
Form &Parameters & $z \epsilon [2.2,5.5)$
& $z \epsilon [2.2,3.0)$ & $z \epsilon [3.0,5.5)$ \\
\tableline
Single & $\log k_1$ &$20.30^{+0.04}_{-0.04}$&$23.51^{+0.08}_{-0.08}$&$19.28^{+0.05}_{-0.05}$\\
& $\alpha_1$ &$-2.04^{+0.09}_{-0.11}$&$-2.21^{+0.19}_{-0.24}$&$-1.99^{+0.10}_{-0.12}$\\
Gamma$^a$ & $\log k_2$ &$-24.36^{+0.04}_{-0.04}$&$-25.15^{+0.08}_{-0.08}$&$-24.04^{+0.05}_{-0.05}$\\
& $\log N_\gamma$ &$>21.7$&$>21.5$&$>21.6$\\
& $\alpha_2$ &$-1.85^{+0.11}_{-0.12}$&$-2.09^{+0.21}_{-0.13}$&$-1.76^{+0.12}_{-0.14}$\\
\tableline
\end{tabular}
\end{center}
\tablenotetext{a}{The $N_\gamma$ parameter reported is a 95$\%$c.l. lower limit.}
\tablecomments{The errors reported are one-parameter errors which do not account for correlations among the parameters.} 
\end{table*}

\section{THE \ion{H}{1} FREQUENCY DISTRIBUTION AND ITS MOMENTS}
\label{sec:fnhi}

\subsection{\fnhi}

\begin{figure}[ht]
\begin{center}
\includegraphics[height=3.5in,angle=90]{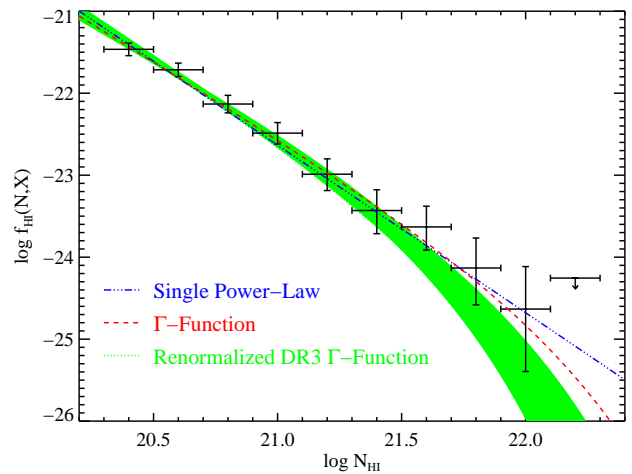}
\end{center}
\caption{\ion{H}{1} frequency distribution \fnhi\ for the PDLA
systems at $z>2.2$ drawn from SDSS-DR5 (binned data points).   
Overplotted on the data are three curves corresponding
to the solutions from maximum likelihood analyses:
(i) a single power-law fit to the PDLA data;
(ii) a $\Gamma$-function fit to the PDLA data;
and
(iii) the $\Gamma$-function fit (with $1\sigma$ uncertainty 
indicated by the green band)
 to the SDSS-DR3 sample of
intervening DLA systems (PHW05) renormalized to match
the incidence of PDLA systems.  
At low \nhi, 
the intervening DLA and PDLA systems exhibit roughly
the same power-law dependence, $\mfnhi \propto \mnhi^{-2}$.
In contrast with the intervening DLA, the PDLA systems
do not show an obvious break in the power-law  
at large \nhi\ values ($> 10^{21.5} \cm{-2}$). 
However,
the difference in \fnhi\ at large
\nhi\ between the intervening DLA (SDSS-DR3) and 
the PDLA systems is only significant at $\approx 95\%$ confidence level.
}
\label{fig:fnhi}
\end{figure}

Akin to the luminosity function of galaxies,
the fundamental measure of a DLA sample is its \ion{H}{1}
frequency distribution, \fnhi.  
Here, $X$ is the absorption distance \citep{bp69} defined
to give a constant comoving distance.
Following PHW05, we
have calculated \fnhi\ for the PDLA systems.  The results
are presented in Figure~\ref{fig:fnhi} and given in
Table~\ref{tab:fnfits}.  We have fitted the frequency distribution
with a single power-law $\mfnhi = k_1 N^{\alpha_1}$
and a $\Gamma$-function \\
$\mfnhi = k_2 \ltp \frac{N}{N_\gamma} \rtp^{\alpha_2} 
\exp \ltp \frac{- N }{N_\gamma} \rtp$.
The maximum likelihood solutions are overplotted in 
the Figure.  The power-law has exponent $\alpha_1 = -2.1 \pm 0.1$ while
the $\Gamma$-function has a similar power-law slope and allows any value 
$\log N_\gamma > 21.6$ at the 95$\%$\,c.l.
We conclude that the PDLA sample does not show a break from
a single power-law model.  
For comparison, we also show the best-fit $\Gamma$-function for
\fnhi\ of intervening DLA from PHW05 renormalized to give
the same number of PDLA systems as observed.  The shaded region
shows the approximate $1\sigma$ uncertainty in the shape of this
$\Gamma$-function.

The figure suggests that the \nhi\ values of the PDLA
sample are not drawn from the same parent population as intervening
DLA systems.
We find, however, that a two-sided Kolmogorov-Smirnov (KS) test 
gives a probability of 30$\%$
for the null hypothesis.  This rather high probability may be misleading,
however, because the standard KS-test focuses on the median of the
two distributions which is dominated by the lower \nhi\ values.
From the Figure, it is evident that the two distributions only 
differ in the high \nhi\ tail of the samples.
Integrating the re-normalized $\Gamma$-function for the intervening
DLA from $\mnhi = 10^{21.5} \cm{-2}$ to infinity, we predict 2.1 absorbers
compared against the 6 that are observed with $\log \mnhi \ge 21.5$.
Even if we ignore the uncertainty in the predicted incidence, the 
Poisson error on six absorbers implies they differ at only $2\sigma$ 
significance (i.e. 95$\%$ c.l.).
Therefore, we contend that the figure is suggestive but there is 
no conclusive evidence for a difference in the shape of \fnhi\ for
the proximate and intervening DLA systems.
We also note that there is no statistically significant
difference in \fnhi\ if we compare two bins of PDLA systems split
at $z=3$ (Table~\ref{tab:fnfits}).

\begin{figure}[ht]
\begin{center}
\includegraphics[width=3.5in]{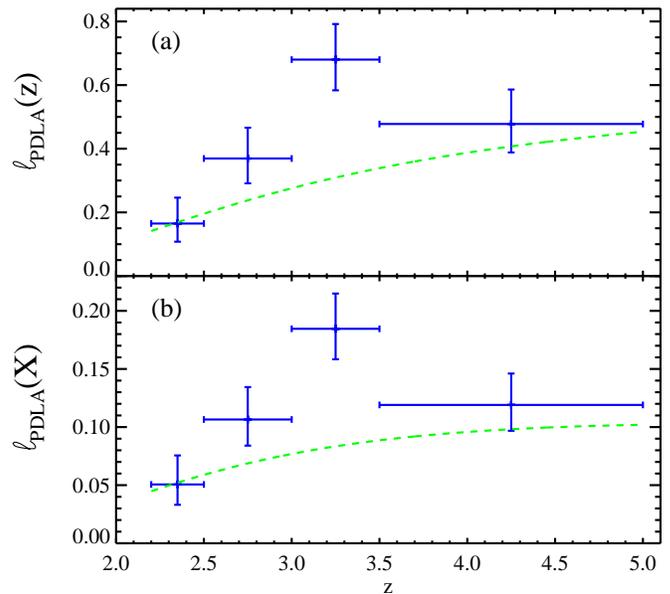}
\end{center}
\caption{Incidence of PDLA systems as a function of redshift
per (a) unit redshift and (b) comoving distance.
The dashed curve traces the incidence of intervening DLA systems
(PHW05; Equation~\ref{eqn:loz}).  
The enhancement in the incidence of PDLA systems at
$z=[2.5,3.5]$ is significant at $>99\%$\,c.l. assuming Poisson
statistics.
}
\label{fig:lozx}
\end{figure}

\begin{table*}[ht]\footnotesize
\begin{center}
\caption{{\sc SUMMARY OF RESULTS\label{tab:summ}}}
\begin{tabular}{lcccccc}
\tableline
\tableline
$z$ & $m_{\rm PDLA}$ & $\Delta z$ & $\ell_{\rm PDLA}(z)$ & $\Delta X$ & 
$\ell_{\rm PDLA}(X)$ & $\Omega_g^{\rm PDLA}$ ($\times 10^{-3}$) \\
\tableline
$\lbrack$2.2,2.5]&   8&  48.5&$0.165^{+0.081}_{-0.057}$& 158.1&$0.051^{+0.025}_{-0.018}$&$ 0.56^{+ 0.23}_{- 0.23}$\\
$\lbrack$2.5,3.0]&  22&  59.6&$0.369^{+0.097}_{-0.078}$& 206.6&$0.107^{+0.028}_{-0.022}$&$ 1.32^{+ 0.46}_{- 0.57}$\\
$\lbrack$3.0,3.5]&  49&  72.0&$0.680^{+0.112}_{-0.097}$& 265.5&$0.185^{+0.030}_{-0.026}$&$ 3.03^{+ 0.74}_{- 0.74}$\\
$\lbrack$3.5,5.0]&  28&  58.6&$0.478^{+0.108}_{-0.090}$& 235.2&$0.119^{+0.027}_{-0.022}$&$ 1.99^{+ 0.67}_{- 0.71}$\\
\tableline
\end{tabular}
\end{center}
\end{table*}

\subsection{THE INCIDENCE OF PDLA SYSTEMS ($\ell_{\rm PDLA}$)}
\label{sec:incidence}
The zeroth moment of the \ion{H}{1} frequency distribution
gives the incidence of absorbers along the sightline. 
Authors traditionally express this quantity per unit redshift,
\loz.  Of greater interest to the nature of the DLA systems
is the incidence per unit absorption distance, \lox.  We will consider
both quantities.   It is standard practice to discretely
estimate \loz\ with a
calculation evaluated over a redshift interval $[z_{min},z_{max}]$,

\begin{equation}
\mloz = \frac{m_{PDLA}}{\smm_{z_{min}}^{z_{max}} g(z)} \cmma
\end{equation}
where $m_{PDLA}$ is the number of DLA systems within the redshift interval.

For our evaluation of \loz\ and \lox\, we include a subtle but important
modification.  Although we have defined the search path of each
quasar to be from $\delta v = 0 \mkms$ to $v_{prox}$ 
(Equations~\ref{eqn:deltaz},\ref{eqn:dv}) for 
quasars with $z_{em} \epsilon [z_{min},z_{max}]$,
we also include all of the absorbers with $\delta v < 0 \mkms$
in our calculation of $m_{PDLA}$.
That is, we increment $m_{PDLA}$ by one for each damped \lya\ system
with $\delta v < 0\mkms$ even though we have defined the
search path $(\Delta z)_j$ to have $\delta v \geq 0 \mkms$.
At first glance, it appears that we will overestimate \loz.
We have adopted this formalism, however, to account for the relatively
large systematic uncertainties in the quasar redshifts
($\S$~\ref{sec:zem}). 
If these errors are roughly symmetric then we will have as many
PDLA shifted outside the sample as shifted in at the boundary 
$v=v_{prox}$.
Similarly, for any given quasar we will be searching a little more or
less redshift path than 3000~\kms\ depending on the sense of the
quasar redshift error, but the total search pathlength will be approximately
correct. At the same time, however, we must include all PDLA systems
with $\delta v < 0 \mkms$ to accurately assess \loz.


Figure~\ref{fig:lozx} presents (a) \loz\
and (b) \lox\ as a function of redshift for the PDLA systems.
The curves in each panel correspond to 

\begin{equation}
<\mloz> = 0.6 \exp(-7 / z^2) 
\label{eqn:loz}
\end{equation}
which is a good representation\footnote{The function is not
properly normalized at $z=0$ but one could add a small offset to 
achieve this.}  
of $<\mloz >$ for the intervening DLA systems at $z>2$.
In fact, a power-law representation of the form
$\mloz \propto (1+z)^\gamma$ \citep[e.g.][]{sw00}
is no longer a good description of the observations at high $z$.

\begin{figure}[ht]
\begin{center}
\includegraphics[height=3.5in,angle=90]{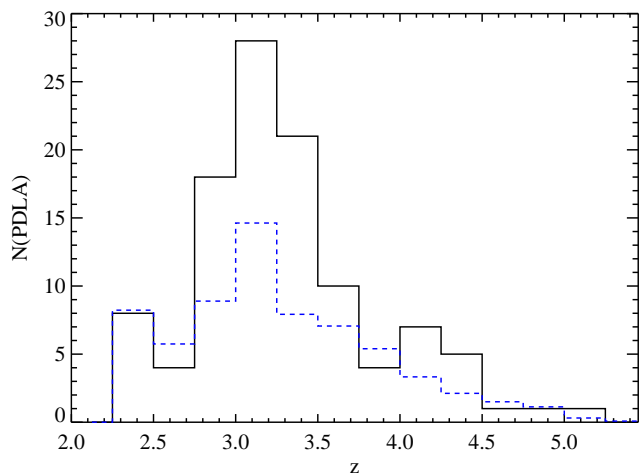}
\end{center}
\caption{Histogram of the absorption redshifts for the full
PDLA sample.  The dashed curve shows the predicted number of 
systems assuming the incidence of intervening DLA systems
(Equation~\ref{eqn:loz}) convolved with the PDLA selection
function (Figure~\ref{fig:gofz}).
}
\label{fig:histlx}
\end{figure}

The two panels show similar results.  One observes an enhancement
in the incidence of PDLA systems for $z = 3$ to 3.5 and that the
incidence of PDLA is consistent with the intervening values at other redshifts.
Because the representation of $\ell$ given by Figure~\ref{fig:lozx}
may be sensitive to the binning, we also compare the proximate
and intervening DLA systems as a function of redshift in histogram
form (Figure~\ref{fig:histlx}).  Here, we have histogrammed the
redshifts of the PDLA systems in the full sample and overplotted the
predicted number calculated by convolving the redshift search path
$g(z)$ with the incidence of intervening DLAs (Equation~\ref{eqn:loz}).

If we assume Poisson statistics, we find that the 95$\%$\,c.l. intervals
of the $z=[2.5,3]$ and $z=[3,3.5]$ do not overlap for the PDLA and
intervening systems.  Similarly, we find that the $99\%$\,c.l. intervals
on $\ell$ for $z=[2.5,3.5]$ do not overlap. 
We conclude, therefore, that the PDLA systems exhibit a higher
incidence than the intervening DLA in the redshift range
$z=[2.5,3.5]$ but are otherwise consistent with the cosmic average.
We will attempt to interpret this rather unusual signal in the
following section.  

\begin{figure}[ht]
\begin{center}
\includegraphics[width=3.5in]{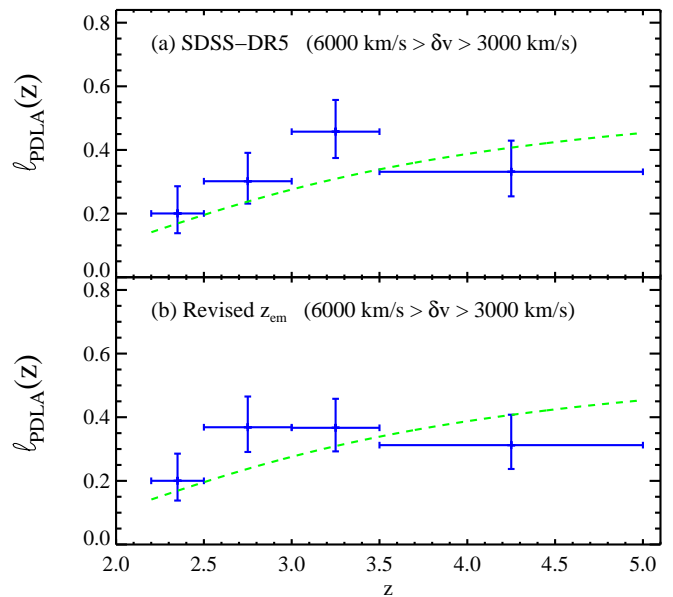}
\end{center}
\caption{
Incidence of DLA systems with $6000 \mkms > \delta v > 3000 \mkms$ as
(a) determined using $z_{em}$ values taken from the SDSS DR5
data release.  Similar to REB06, we note an enhancement in
\lpdla\ at $z \sim 3$.
After recalculating $z_{em}$ for all quasars with DLA
candidates within 10,000\kms, we derive the results shown
in panel (b).  The incidence of DLA in this velocity interval
is consistent ($\chi^2_nu = 1.04$ assuming four degrees of freedom)
with the expected rate assuming the cosmic average with no error in this value.
}
\label{fig:loz6000}
\end{figure}

A puzzling result from REB06 was that the authors reported an
enhancement in \loz\ for DLA systems with $3000 \mkms < \delta v <
6000 \mkms$, which even exceeded that for PDLAs with $\delta v < 3000
\mkms$.  Such an enhancement is very unlikely to be related to
absorber clustering near quasars.  At $z \approx 3$, $\delta v = 3000
\mkms$ corresponds to $\approx 30 {\rm Mpc} \, h^{-1}$.  Even for the
strong high redshift quasar clustering measured by \citet{shen07}
(e.g.\ $\xi(r) = (r/r_0)^{-\gamma}$ with $r_0 \sim 17 {\rm Mpc} \,
h^{-1}$ and $\gamma=2$ at $z\gtrsim 3)$ a relatively small enhancement
is expected at such large distances from the quasar.  In
Figure~\ref{fig:loz6000} we present \loz\ for $3000 \mkms < \delta v <
6000 \mkms$ adopting (a) the SDSS-DR5 quasar redshifts and (b) our
revised values.  Although the latter shows a mild enhancement in \loz\ at
$z\approx 3$, the effect is significantly smaller than REB06.
Adopting the revised redshifts, furthermore, we derive \loz\ values
consistent with the cosmic average.  We suspect, therefore, that the
REB06 result for $v_{prox} = 6000\mkms$ was predominantly the result of
spurious redshifts,
although they reached qualitatively similar conclusions as our own
when restricting to $\delta v < 3000 \mkms$.

REB06 also reported an enhancement in PDLA of a factor 
$\mloz = 1.4 <\mloz>$ with $\approx 2 \sigma$ significance
at a mean redshift $<z_{abs}> = 3.36$.
Our results indicate a larger enhancement and much
higher statistical significance at this redshift.
Again, the 
differences are related to our larger sample size and 
errors in the $z_{em}$ values reported by the SDSS.
We also note that our results more resemble those from \cite{eyh+02}  
who had more precisely measured QSO redshifts given the lower 
median redshift of their sample.


\begin{figure}[ht]
\begin{center}
\includegraphics[height=3.5in,angle=90]{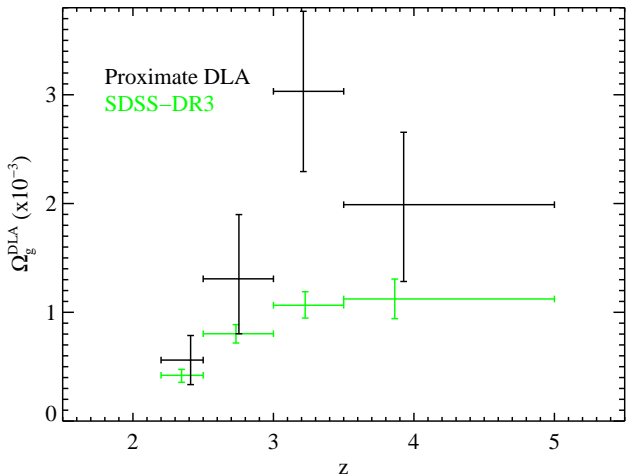}
\end{center}
\caption{
Mass density in neutral gas relative to the critical density for
intervening DLA (light green points; PHW05) and the
PDLAs (black).  Similar to the results on the incidence of PDLA,
we observe an enhancement in \oprox\ at $z \sim 3$ yet results
consistent with the cosmic average at other redshifts.
}
\label{fig:omg}
\end{figure}

\subsection{THE \ion{H}{1} MASS DENSITY WITHIN $\approx 30 h^{-1}$\,Mpc OF QUASARS (\oprox)}
\label{sec:omg}

The first moment of the \ion{H}{1} frequency distribution yields the 
mass density in \ion{H}{1} atoms:

\begin{equation}
\Omega_g(X) dX \equiv \frac{\mu m_H H_0}{c \rho_c} 
  \intl_{N_{min}}^{N_{max}} N \mfnhi dX \perd
\label{eqn:omg}
\end{equation}
\noindent where $\mu$ is the mean particle mass per $m_H$ of the gas 
(taken to be 1.3), $H_0$ is Hubble's constant, and $\rho_c$ is the critical mass
density.  Because the damped \lya\ systems dominate this integral
\citep{ph04,opb+07} and because they are predominantly neutral, an
evaluation of Equation~\ref{eqn:omg} from $N_{min} = 2 \sci{20}
\cm{-2}$ to infinity gives an accurate estimate of the mass density of
neutral gas.  For this analysis, we have focused on proximate DLA
systems with $\delta v \leq 3000 \mkms$ roughly correspondingly to a
comoving 30\,Mpc~$h^{-1}$.  We will refer to this quantity as \oprox,
the average mass density of neutral gas measured along an 
$\approx 30 {\rm Mpc}~h^{-1}$ path toward quasars with a median
$i$-magnitude of 19.1\,mag.  

Following standard practice (e.g.\ PHW05), we estimate \oprox\ with a discrete
evaluation:

\begin{equation}
\Omega_g^{\rm PDLA} = \frac{\mu m_H H_0}{c \rho_c} \frac{\sum N_{\rm HI}}{\Delta X} \;\; ,
\label{eqn:omgdisc}
\end{equation}
where the sum is performed over the \nhi\ measurements
of the damped \lya\ systems in a given redshift interval with 
survey pathlength $\Delta X$.
The results are presented in Figure~\ref{fig:omg} and compared against the
mass density for intervening damped \lya\ systems, \odla,
determined from the DR3 sample (PHW05). 
Similar to the zeroth moment of \fnhi, we find that \oprox\ is consistent
with \odla\ at $z<3$ but exceeds \odla\ at $z \approx 3.5$.  
Although the uncertainty in \oprox\ is large, the enhancement over \odla\ 
is significant at greater than $95\%$ c.l. in the $z=3$ to 3.5 interval.

There are two additional points to emphasize with regards to our 
estimation of \oprox.  First, we do not have a large enough sample of
proximate DLAs to determine the inevitable break in the $\alpha \approx -2$
power-law for \fnhi\ observed for $\mnhi < 10^{21.3} \cm{-2}$.
Therefore, our evaluation of \oprox\ is formally a lower limit at
all redshifts.
Second, as noted above, we are measuring \oprox\ along biased
sightlines.  If bright quasars photoionize significant quantities of
neutral hydrogen gas and anisotropically, we will underestimate \oprox\
if magnitude-limited quasar surveys are biased to these viewing angles.
We suspect that the first effect (sample size) is a less than factor of
two effect while we will consider simple models of the latter effect
in the following section.

We have performed the following, simple calcluation in an attempt
to estimate a lower
limit to the dark matter mass of halos hosting quasars.
At $z=3$, we have calculated the dark matter density within
30\,Mpc\,$h^{-1}$ of the most massive dark matter halos in a 
cosmological box 600\,Mpc\,$h^{-1}$ on a side.
The mean value never exceeds 1.5 times the mean density at this epoch.
We reach the obvious conclusion that the \ion{H}{1} mass density 
must increase non-linearly (e.g.\ $n_H^2$) with dark matter overdensity.
Therefore, a direct comparison with our observations will 
require a full hydronamical simulation that includes an
extragalactic background radiation field and radiative
transfer.

\section{DISCUSSION}
\label{sec:discuss}
In this section we physically interpret the incidence of PDLA systems
measured in \S~\ref{sec:incidence}, by considering two complimentary
measurements: the clustering of optically thick absorbers around
quasars in the \emph{transverse} direction, measured from close
projected pairs of quasars \citep{hp06}, and the strength and
evolution of the auto-correlation function of high-redshift quasars
\citep{shen07}.  Because PDLAs can be detected in absorption against
bright background quasars out to $z\sim 5$ their statistics probe the
environments of quasars to much higher redshifts\footnote{This
  redshift limit is set by the small number of quasars with $z>5$.}
than galaxy cross-correlation studies \citep{as05,coil06}. However,
the large ionizing flux emitted by a quasar can ionize the neutral gas
in nearby galaxies, thus reducing their absorption cross-section
\citep{hp06}.  The interpretation of PDLA incidence as a measure of
quasar-galaxy clustering is thus complicated by ionization effects.
Nevertheless, the physical problem of a quasar illuminating a
self-shielding optically thick absorber is very rich and can teach us
much about both quasars and high-redshift galaxies.

\begin{figure}[ht]
\plotone{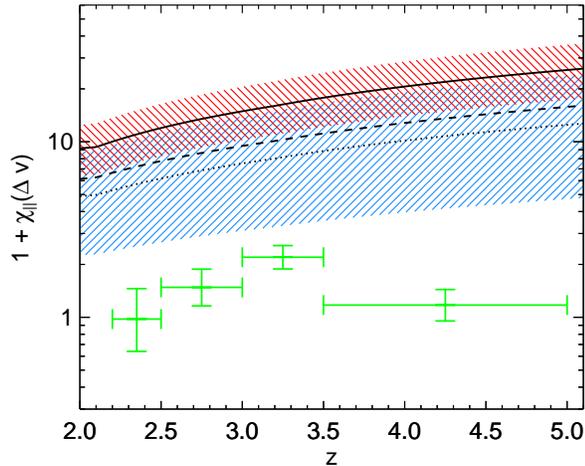}
\caption{Comparison of the evolution of the overdensity of PDLAs near
  quasars $(1 +\chi_{\parallel})$ (see eqn.~\ref{eqn:xi_cross}) to
  those implied by other measurements.  The (green) data points show
  our measurements of the line-of-sight correlation function from the
  abundance of PDLAs, where the horizontal error bars indicate the
  size of the redshift bins used.  The solid curve shows the
  line-of-sight correlation function prediction obtained by combining
  the \citet{shen07} clustering results with the \citet{hp06}
  transverse quasar-absorber clustering measurement for
  $\gamma=2$. The upper (red) striped region illustrates $1-\sigma$
  error from combining these results.  The dashed curve shows the same
  quantity, but using the \citet{hp06} measurement with a shallower
  slope of $\gamma=1.6$ (the $1-\sigma$ error region is comparable to
  the red striped region but not shown for clarity). The dotted curve
  and lower (blue) striped region are the line-of-sight correlation
  function and corresponding $1-\sigma$ error when we combine the
  \citet{shen07} clustering results with DLA auto-correlation implied
  by the \citet{cwg+06-2} measurements. The (red-blue) cross-hatching
  indicates the overlap of the error regions, and illustrates that the
  \citep{hp06} results and \citet{cwg+06-2} measurement are consistent
  within the errors.}
\label{fig:model}
\end{figure}

Following \citet{hp06}, we can describe the increase or decrease in the line
density, \loz, of proximate DLAs at a velocity separation $v$ from 
a quasar with the `line-of-sight' (LOS) correlation function 
\begin{equation}
  \ell(z,v,\Delta v)
=\left\langle\ell(z)\right\rangle [1+ \chi_{\parallel}(v,\Delta
  v)],\label{eqn:los} 
\end{equation}
where $\langle\ell(z)\rangle$ represents the cosmic average
line density (PHW05) and $\Delta v$ indicates the 
width of the velocity interval searched. The LOS correlation function 
is given by an average of the quasar-absorber correlation function, $\xi_{\rm
  QSO-DLA}(r)$, over a cylinder with cross-section $A$ equal to the
DLA cross-section and extent $2\Delta v \slash a H(z)$ along the
line-of-sight direction.  Here $a$ is the scale factor and $H(z)$ is
the Hubble constant at redshift $z$. For velocity separations $v$
close to the quasar, $\chi_{\parallel}$ depends on the size of the
absorber cross-section, which is unknown. In what follows, we assume a
plausible radius of $r_{abs} \equiv \sqrt{A\slash \pi} = 20~{\rm kpc}$.
Increasing $r_{abs}$ by a factor of two results in a factor of two
decrease in $\chi_{\parallel}$ \citep[see Figure~2 of][]{hp06}.

Using projected pairs of quasars, \citet{hp06} measured the
cross-correlation between foreground quasars at $z\sim 2.5$ and
optically thick $\mnhi > 10^{19}~{\rm cm}^2$ absorbers, detected in
the background quasar spectra.  Assuming a power law shape for the
quasar-absorber cross-correlation function, they measured $r_0 =
9.2^{+1.5}_{-1.7}~\hMpc$ for $\gamma=1.6$, or $r_0 =
5.8^{+1.0}_{-0.6}~\hMpc$ for $\gamma=2$. This measurement of the
\emph{transverse} clustering of absorbers around quasars should be
commensurate with the line-of-sight clustering probed by the incidence
of PDLA systems, provided that the clustering pattern of absorbers
around quasars is isotropic. However, \citet{hp06} argued that the
strength of the transverse clustering predicts that $\sim 15-50\%$ of
\emph{all} quasars should show a $\mnhi > 10^{19}~{\rm cm}^2$ absorber
within $\Delta v < 3000$~\kms, which is not observed \citep{opb+07}.  Thus,
the transverse clustering overpredicts the number of absorbers along
the line-of-sight by a large factor, or equivalently the clustering
pattern is highly anisotropic. The most plausible physical explanation
is that the transverse direction is less likely to be illuminated by
ionizing photons than the line-of-sight, either due to anisotropic
emission, or variability of the quasar emission on timescales short
compared to the transverse light crossing time ($\sim 5\times
10^5~{\rm yr}$).  We revisit this issue of anisotropic clustering with
the statistics of PDLAs below.

\citet{shen07} recently quantified the clustering of high-redshift
($z\gtrsim 2.9$) quasars and measured a comoving auto-correlation length
of $r_0=16.9\pm 1.7~\hMpc$ for quasars in the redshift range $2.9 \le
z \le 3.5$, and an even stronger $r_0=24.3\pm 2.4~\hMpc$ for $z\ge
3.5$ quasars. High redshift quasars are thus much more strongly
clustered than their $z\sim 2$ counterparts, which have $r_0\approx
7.5~\hMpc$ \citep[see e.g.][]{pmn04,croom05,pn06}, and the evolution
of the quasar clustering strength with redshift is extremely
rapid. Naively, we expect this to drive a rapid
evolution in the cross-correlation of quasars with DLAs.


Our strategy for comparing the PDLA clustering to these complimentary
measurements is to normalize the quasar-DLA
cross-correlation strength to the transverse \citet{hp06} result at
$z\sim 2.5$, and assume that the clustering of DLAs does not evolve
with redshift, but that the quasar clustering evolves as measured by
\citet{shen07}. Specifically, we use the relation
\begin{equation}
\xi_{\rm
  QSO-DLA}(r,z) = \sqrt{\xi_{\rm QSO-QSO}(r,z)\xi_{\rm DLA-DLA}(r)} 
\label{eqn:xi_cross}, 
\end{equation}
to determine the parameters of the $\xi_{\rm DLA-DLA}$ and propagate
errors according to those quoted by both measurements. \citet{shen07}
measured the auto-correlation function of quasars in two redshift bins
$2.9 < z < 3.5$ and $z > 3.5$, very similar to the two high redshift
bins in this work.  Quasar clustering has not yet been measured in 
the redshift range $2.1 < z < 2.9$, so we linearly interpolate between 
the \citet{shen07} result and the redshift evolution measured by
\citet{pn06}\footnote{ A small correction is applied to account for
  the fact that Shen et al. used a different power law slope
  ($\gamma=2$) than Porciani \& Nordberg ($\gamma=1.8$).}.

A few limitations to our approach should be noted. First, by assuming
the DLA clustering does not evolve, we might overestimate
(underestimate) the strength of the quasar-DLA cross-correlation
function if DLA-galaxies are less (more) clustered at high
redshift. However, \citet{adel05} found that the clustering of LBGs is
very nearly constant in the range $r_0 \sim 4-5~\hMpc$ over the
redshift range $1.4 \le z \le 3.5$, and \citet{Ouchi04} similarly
measured the correlation lengths of $3.5 \le z \le 5.2$ LBGs and
found a correlation length of $r_0\sim 5~\hMpc$ which was nearly
constant with redshift. In light of evidence that DLA-galaxies are
drawn from a similar (but fainter) galaxy population as 
spectroscopically selected LBGs
\citep{mwf+02,schaye01}, it is reasonable to assume that DLA
clustering does not evolve significantly.  Second, the \cite{hp06}
measurement was for all absorbers with $\mnhi > 10^{19}~{\rm cm}^2$,
so by applying that measurement to the DLAs ($\mnhi \ge 2\times
10^{20}~\cm{-2}$) we implicitly assume that the clustering of
absorbers is independent of column density\footnote{This is a
  reasonable assumption if absorbers with $\mnhi \approx 10^{19} \cm{-2}$
arise from material at larger impact parameters from DLA systems
\citep[e.g.][]{mps+03}.}. Finally, it is conceivable that
the \citep{hp06} cross-correlation strength is over-estimated because
of line-blending of systems near the column density threshold ($\mnhi
> 10^{19}~\cm{-2}$), which results in a `Malmquist'-type bias
because the line density \loz\ of absorbers is a relatively 
steep function \citep{opb+07} of column density limit. 

To address these issues we explore another avenue for estimating the
autocorrelation of $\xi_{\rm DLA-DLA}$ in
eqn.~\ref{eqn:xi_cross}. Namely, \citet{cwg+06-2} used photometrically
selected LBGs in the vicinity of DLAs to measure the cross-correlation
between LBGs and DLAs at $z\sim 3$. Assuming $\gamma=1.6$ they
measured $r_0=2.93^{+1.4}_{-1.5}$ for the DLA-LBG cross-correlation,
and $r_0=3.32\pm 0.6$ for the LBG auto-correlation.  From
$\xi_{\rm DLA-DLA} = \sqrt{\xi_{\rm DLA-LBG}\slash \xi_{\rm
    LBG-LBG}}$, we deduce the auto-correlation of DLAs to be $r_0 =
2.6\pm 2.7~\hMpc$ ($\gamma=1.6$), which can then be used in
eqn.~\ref{eqn:xi_cross}, again assuming that that the DLA clustering
does not evolve with redshift.

  
Our measurement of the evolution of the overdensity of PDLAs near
quasars $(1 + \chi_{\parallel})$ is compared to the implications of
these other measurements in Figure~\ref{fig:model}.  The (green) data
points illustrate our measurement of the line-of-sight correlation
function from the abundance of PDLAs (see eqn.~\ref{eqn:los}). The
solid curve shows the line-of-sight correlation function prediction
when we combine the \citet{shen07} clustering results with the
\citet{hp06} quasar-absorber clustering measurement for
$\gamma=2$. The upper (red) striped region illustrates the implied
$1\sigma$ error from combining these results.  The dashed curve shows
the same quantity, but using the \citet{hp06} measurement with a
shallower slope of $\gamma=1.6$ (the $1\sigma$ error region is
comparable to the red striped region but not shown for clarity). The
dotted curve and lower (blue) striped region are the line-of-sight
correlation function and corresponding $1\sigma$ error when we
combine the \citet{shen07} clustering results with DLA
auto-correlation implied by the \citet{cwg+06-2} measurements. The
(red-blue) cross-hatching indicates the overlap of the error regions,
and illustrates that the \citep{hp06} results and \citet{cwg+06-2}
measurement are consistent within the errors.

A discrepancy is apparent between the predictions in
Figure~\ref{fig:model}, which are driven by the strength of the quasar
auto-correlation function and its rapid redshift evolution, and the
relatively modest clustering of PDLAs around quasars and lack of
evolution, measured here. This disagreement provides compelling
evidence for the hypothesis that the ionizing flux from quasars
photoevaporates \ion{H}{1} in nearby DLA-galaxies, thus reducing their
cross-section for DLA absorption. \citet{hp06} used a simple
photoevaporation model to show that optically thick absorbers with
$n_{\rm H} \lesssim 0.1 \cm{-3}$ will be photoevaporated if they lie within
$\sim 1~{\rm Mpc}$ of a luminous quasar, so our results are sensible
on physical grounds.


The decrease in the number of optically thin \lya\ forest absorption
lines in the vicinity of quasars, known as the \emph{proximity
  effect}, has been detected and is well studied
\citep{bdo88,sbd+00,claude07}. The factor of $\sim 5-10$ discrepancy
between the predicted clustering results and the data in
Figure~\ref{fig:model} provides compelling evidence that a similar
proximity effect exists for optically thick absorption line systems.
However, according to \citet{hp06}, there is no transverse proximity
effect for optically thick absorbers, which gains further credibility
in light of the null detections of the transverse effect in the
(optically thin) Ly$\alpha$ forest \citep[][but see Worseck \&
Wisotzki (2006)]{Crotts89,DB91,FS95,LW01,Schirber04,Croft04}, although
these transverse studies are all based on only a handful of projected
pairs.  Both the optically thin results and our result for optically
thick systems can be explained if the transverse direction is less
likely to be illuminated by ionizing photons than the line-of-sight,
either because the emission is anisotropic, or because the foreground
quasar varies on timescales short compared to the transverse light
crossing time ($\sim 5\times 10^5~{\rm yr}$) \citep[see][for a
detailed discussion of both]{hp06}.

\begin{figure}[ht]
\plotone{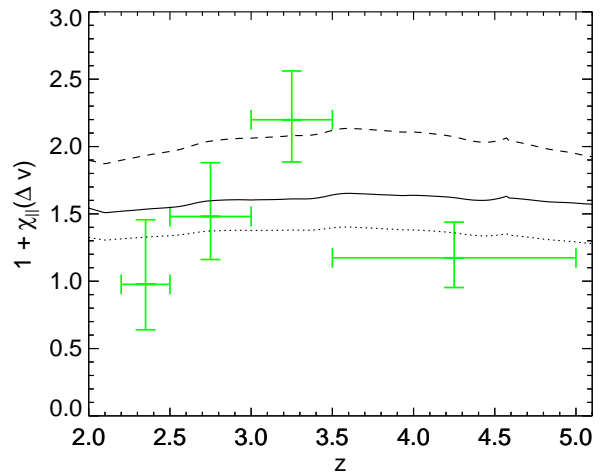}
\caption{Toy photoevaporation model predictions for the evolution of
  the overdensity of PDLAs near quasars. The solid, dashed, and dotted
  curves represent the same measurements shown in
  Figure~\ref{fig:model} but allowing for a reduction in the
  clustering signal because optically thick absorbers are
  photoevaporated by the large ionizing flux if they are close
  enough to the quasar. The volume density of neutral hydrogen was assumed
  to be $n_{\rm H}= 0.1~{\rm cm}^3$.}
\label{fig:evap}
\end{figure}

We use a toy model to illustrate how quasar-absorber clustering can be
used to constrain the physical properties of DLAs following
\citet{hp06}, who introduced a photoevaporation criterion for
optically thick absorbers illuminated by quasars, motivated by the
work of \citet{bertoldi89}. This criterion allows us to compute a
minimum distance from the quasar, as a function of volume density, at
which a DLA can self-shield against photoevaporation. In this context,
the curves in Figure~\ref{fig:model} represent predictions for the
\emph{intrinsic} clustering of DLAs around quasars in the absence of
ionization effects. Because proximate absorbers lie along the
line-of-sight, they must be exposed to the quasars ionizing flux, and
we can calculate the reduction in the clustering strength because of
photoevaporation \citep[see][for details]{hp06}.  The solid, dashed,
and dotted curves in Figure~\ref{fig:evap} represent predictions from
the same measurements shown in Figure~\ref{fig:model}, but after
taking photoevaporation into account using this simple approach. We
used a volume density of hydrogen of $n_{\rm H} = 0.1~{\rm cm}^{-3}$
and assumed the quasar had an average $i$-band magnitude $i=19.1$,
which is the median of our PDLA sample.  Although crude, this simple
model illustrates how the density distribution in DLAs can be measured
by comparing the abundance of PDLAs to the intrinsic quasar-DLA
clustering, deduced either from the transverse quasar-DLA clustering
or by combining independent measurements of quasar and DLA clustering.


\section{SUMMARY}
\label{sec:summary}

We surveyed the spectra of 5938 SDSS quasars in the redshift range
$2.2\lesssim z \lesssim 5$ for proximate DLAs within 3000~\kms\ of the
quasar emission redshift, and presented the largest sample (\npdla\
systems) of PDLAs uncovered to date. Robust systemic redshifts of the
quasars hosting these PDLAs were computed using a redshift estimator
which which accounts for the relative shifts between quasar emission
lines and the systemic frame, and which ignores the \lya\ emission
line, which can be entirely absorbed for quasars with PDLAs. These
improved redshifts allowed us to measure the abundance and distribution
of PDLAs near quasars. The primary conclusions of this study are:

\begin{enumerate}

\item  There is suggestive evidence that the \nhi\ values of the PDLA
  population are not drawn from the same parent population as intervening
  DLA systems. Specifically, the PDLA sample does not show evidence for 
  a break from a single power-law model.  However, the statistical significance 
  of the difference between the PDLA column density distribution and 
  the underlying distribution of intervening DLAs is only at the 
  $2\sigma$ level (i.e. 95$\%$ c.l.).
  
\item PDLA systems exhibit a higher incidence than the intervening
  DLAs in the redshift range $z\epsilon [2.5,3.5]$, but are consistent
  with the cosmic average at lower ($z \epsilon [2.2,2.5]$) and higher
  ($z \epsilon [3.5,5.0]$) redshifts.  

\item The average mass density meaured along an
  $\approx 30 {\rm Mpc} h^{-1}$ path toward the quasars we
  surveyed, \oprox, is consistent with the cosmic average, \odla, at
  $z<3.0$ and $z>3.5$. In the redshift interval $z\epsilon [3.0,3.5]$
  an enhancement in \oprox is detected at greater than  $95\%$ c.l.
  with central value \oprox\=  2 to 3 times \odla.

\item A comparison of the strength of the quasar auto-correlation
  function and its rapid redshift evolution, to the relatively modest
  clustering of PDLAs around quasars and a lack of significant
  evolution, provides compelling evidence for the hypothesis that the
  ionizing flux from quasars photoevaporates \ion{H}{1} in nearby
  DLA-galaxies, thus reducing their cross-section for DLA absorption
  \citep{hp06}. 
\end{enumerate}




\acknowledgments

We are grateful to S. Ellison for helpful discussions and for
carefully reading an early version of this manuscript. We thank
J. Diemond for performing the numerical experiment described in
$\S$~\ref{sec:omg}.  J.X.P. is partially supported by NSF CAREER grant
(AST-0548180).  JFH is supported by NASA through Hubble Fellowship
grant \# 01172.01-A, awarded by the Space Telescope Science Institute,
which is operated by the Association of Universities for Research in
Astronomy, Inc., for NASA, under contract NAS 5-26555.




\end{document}